\documentclass[aps,prl,twocolumn,showpacs,preprintnumbers,amsmath,amssymb,refmerge,superscriptaddress]{revtex4}
\usepackage{color}
\usepackage{graphicx}
\usepackage{dcolumn}
\usepackage{longtable}
\newcommand{\ra}{\rightarrow}
\newcommand{\lam}{\Lambda}

\newcommand{\xim}{\Xi^-}

\begin{document}


\preprint{\vbox{ \hbox{   }
%
							   \hbox{Belle Preprint 2013-13} 
							   \hbox{KEK Preprint 2013-23}
}}

\title{First observation of Cabibbo-suppressed $\Xi_c^0$ decays}

\noaffiliation
\affiliation{University of the Basque Country UPV/EHU, 48080 Bilbao}
\affiliation{University of Bonn, 53115 Bonn}
\affiliation{Budker Institute of Nuclear Physics SB RAS and Novosibirsk State University, Novosibirsk 630090}
\affiliation{Faculty of Mathematics and Physics, Charles University, 121 16 Prague}
\affiliation{University of Cincinnati, Cincinnati, Ohio 45221}
\affiliation{Deutsches Elektronen-Synchrotron, 22607 Hamburg}
\affiliation{Justus-Liebig-Universit\"at Gie\ss{}en, 35392 Gie\ss{}en}
\affiliation{Gifu University, Gifu 501-1193}
\affiliation{II. Physikalisches Institut, Georg-August-Universit\"at G\"ottingen, 37073 G\"ottingen}
\affiliation{Hanyang University, Seoul 133-791}
\affiliation{University of Hawaii, Honolulu, Hawaii 96822}
\affiliation{High Energy Accelerator Research Organization (KEK), Tsukuba 305-0801}
\affiliation{Ikerbasque, 48011 Bilbao}
\affiliation{Indian Institute of Technology Guwahati, Assam 781039}
\affiliation{Indian Institute of Technology Madras, Chennai 600036}
\affiliation{Institute of High Energy Physics, Chinese Academy of Sciences, Beijing 100049}
\affiliation{Institute of High Energy Physics, Vienna 1050}
\affiliation{Institute for High Energy Physics, Protvino 142281}
\affiliation{INFN - Sezione di Torino, 10125 Torino}
\affiliation{Institute for Theoretical and Experimental Physics, Moscow 117218}
\affiliation{J. Stefan Institute, 1000 Ljubljana}
\affiliation{Kanagawa University, Yokohama 221-8686}
\affiliation{Institut f\"ur Experimentelle Kernphysik, Karlsruher Institut f\"ur Technologie, 76131 Karlsruhe}
\affiliation{Korea Institute of Science and Technology Information, Daejeon 305-806}
\affiliation{Korea University, Seoul 136-713}
\affiliation{Kyungpook National University, Daegu 702-701}
\affiliation{\'Ecole Polytechnique F\'ed\'erale de Lausanne (EPFL), Lausanne 1015}
\affiliation{Faculty of Mathematics and Physics, University of Ljubljana, 1000 Ljubljana}
\affiliation{University of Maribor, 2000 Maribor}
\affiliation{Max-Planck-Institut f\"ur Physik, 80805 M\"unchen}
\affiliation{School of Physics, University of Melbourne, Victoria 3010}
\affiliation{Moscow Physical Engineering Institute, Moscow 115409}
\affiliation{Moscow Institute of Physics and Technology, Moscow Region 141700}
\affiliation{Graduate School of Science, Nagoya University, Nagoya 464-8602}
\affiliation{Kobayashi-Maskawa Institute, Nagoya University, Nagoya 464-8602}
\affiliation{Nara Women's University, Nara 630-8506}
\affiliation{National Central University, Chung-li 32054}
\affiliation{National United University, Miao Li 36003}
\affiliation{Department of Physics, National Taiwan University, Taipei 10617}
\affiliation{H. Niewodniczanski Institute of Nuclear Physics, Krakow 31-342}
\affiliation{Nippon Dental University, Niigata 951-8580}
\affiliation{Niigata University, Niigata 950-2181}
\affiliation{Osaka City University, Osaka 558-8585}
\affiliation{Pacific Northwest National Laboratory, Richland, Washington 99352}
\affiliation{Panjab University, Chandigarh 160014}
\affiliation{University of Pittsburgh, Pittsburgh, Pennsylvania 15260}
\affiliation{Research Center for Electron Photon Science, Tohoku University, Sendai 980-8578}
\affiliation{University of Science and Technology of China, Hefei 230026}
\affiliation{Seoul National University, Seoul 151-742}
\affiliation{Soongsil University, Seoul 156-743}
\affiliation{Sungkyunkwan University, Suwon 440-746}
\affiliation{School of Physics, University of Sydney, New South Wales 2006}
\affiliation{Tata Institute of Fundamental Research, Mumbai 400005}
\affiliation{Excellence Cluster Universe, Technische Universit\"at M\"unchen, 85748 Garching}
\affiliation{Toho University, Funabashi 274-8510}
\affiliation{Tohoku Gakuin University, Tagajo 985-8537}
\affiliation{Tohoku University, Sendai 980-8578}
\affiliation{Department of Physics, University of Tokyo, Tokyo 113-0033}
\affiliation{Tokyo Institute of Technology, Tokyo 152-8550}
\affiliation{Tokyo Metropolitan University, Tokyo 192-0397}
\affiliation{Tokyo University of Agriculture and Technology, Tokyo 184-8588}
\affiliation{University of Torino, 10124 Torino}
\affiliation{CNP, Virginia Polytechnic Institute and State University, Blacksburg, Virginia 24061}
\affiliation{Wayne State University, Detroit, Michigan 48202}
\affiliation{Yamagata University, Yamagata 990-8560}
\affiliation{Yonsei University, Seoul 120-749}
 \author{R.~Chistov}\affiliation{Institute for Theoretical and Experimental Physics, Moscow 117218} 
  \author{I.~Adachi}\affiliation{High Energy Accelerator Research Organization (KEK), Tsukuba 305-0801} 
  \author{H.~Aihara}\affiliation{Department of Physics, University of Tokyo, Tokyo 113-0033} 
  \author{D.~M.~Asner}\affiliation{Pacific Northwest National Laboratory, Richland, Washington 99352} 
  \author{V.~Aulchenko}\affiliation{Budker Institute of Nuclear Physics SB RAS and Novosibirsk State University, Novosibirsk 630090} 
  \author{T.~Aushev}\affiliation{Institute for Theoretical and Experimental Physics, Moscow 117218} 
  \author{A.~M.~Bakich}\affiliation{School of Physics, University of Sydney, NSW 2006} 
  \author{A.~Bala}\affiliation{Panjab University, Chandigarh 160014} 
  \author{V.~Bhardwaj}\affiliation{Nara Women's University, Nara 630-8506} 
  \author{B.~Bhuyan}\affiliation{Indian Institute of Technology Guwahati, Assam 781039} 
% ...
\collaboration{The Belle Collaboration}



\begin{abstract}

We report the first observation of the Cabibbo-suppressed decays 
$\Xi_c^0\rightarrow \Xi^- K^+$, $\Xi_c^0\ra\Lambda K^+ K^-$ 
 and $\Xi_c^0\rightarrow \Lambda \phi$,  
using a data sample of
$711\,~\mathrm{fb}^{-1}$ collected at the $\Upsilon(4S)$ resonance  
with the Belle detector at the KEKB asymmetric-energy $e^+ e^-$  collider. 
We measure the ratios of branching fractions to be    
$\frac{{\cal B}(\Xi_c^0\ra \Xi^- K^+)}{{\cal B}(\Xi^0_c\ra\Xi^-\pi^+)}=(2.75\pm 0.51\pm 0.25)\times 10^{-2}$, 
$\frac{{\cal B}(\Xi_c^0\ra\Lambda K^+ K^-)}{{\cal B}(\Xi_c^0\rightarrow\Xi^- \pi^+)}=(2.86\pm 0.61\pm 0.37)\times 10^{-2}$  and
$\frac{{\cal B}(\Xi_c^0\ra \Lambda\phi)}{{\cal B}(\Xi^0_c\ra\Xi^-\pi^+)}=(3.43\pm 0.58\pm 0.32)\times 10^{-2}$, 
where the first uncertainty is statistical and the second is systematic.

\end{abstract}


\maketitle

{\renewcommand{\thefootnote}{\fnsymbol{footnote}}}
\setcounter{footnote}{0}

\begin{figure}[!h]
\includegraphics[width=0.25\textwidth]{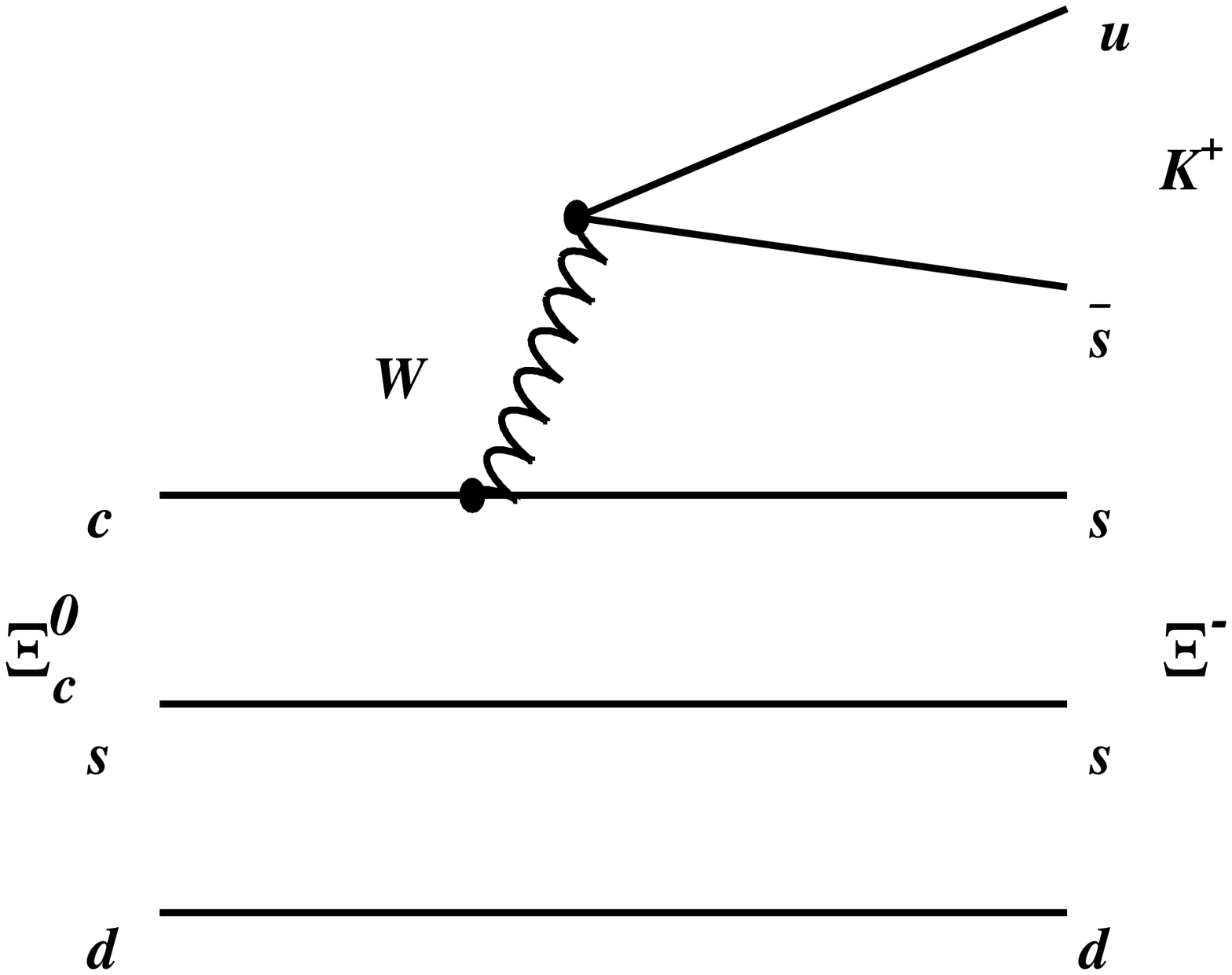}
\includegraphics[width=0.25\textwidth]{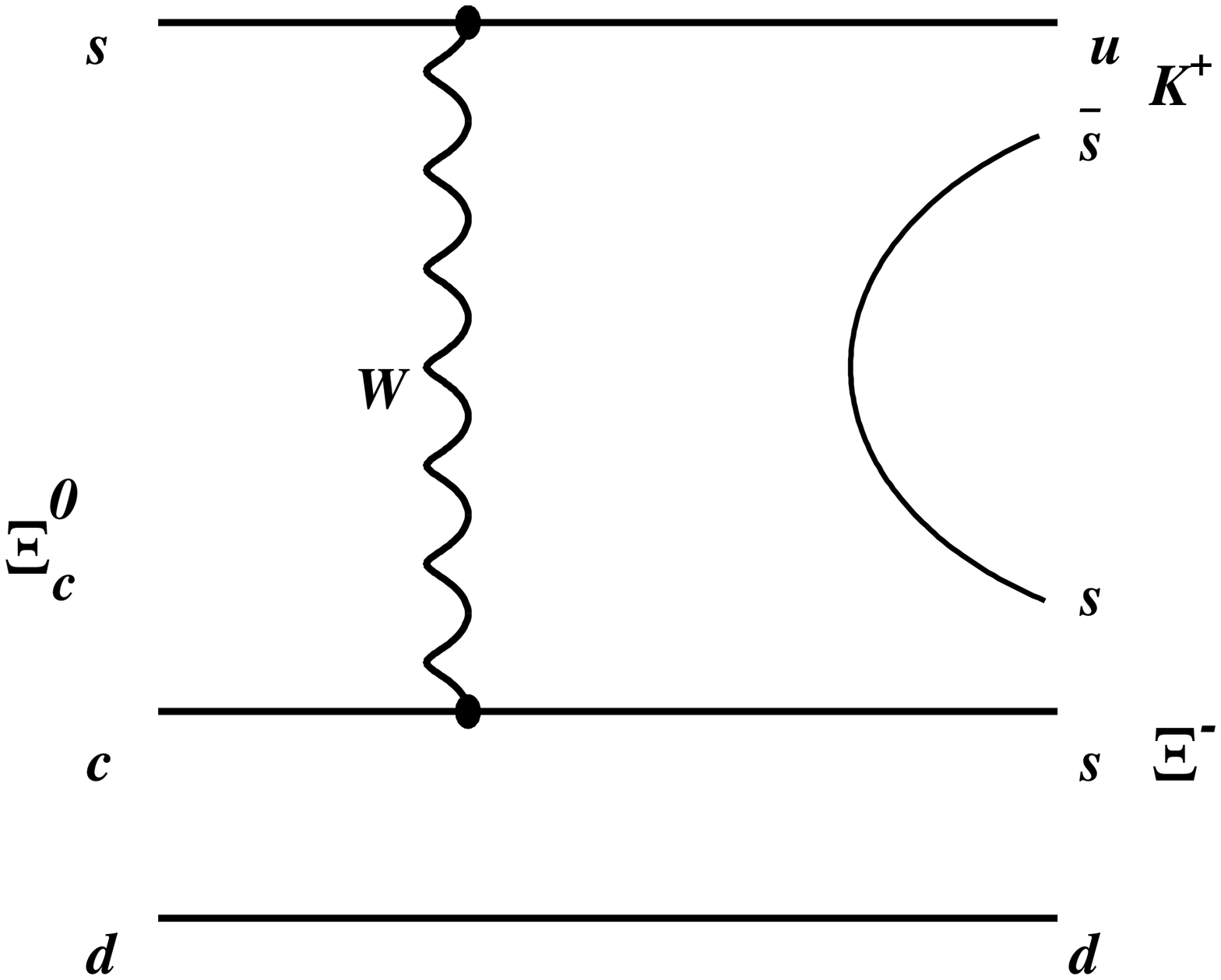}
\includegraphics[width=0.25\textwidth]{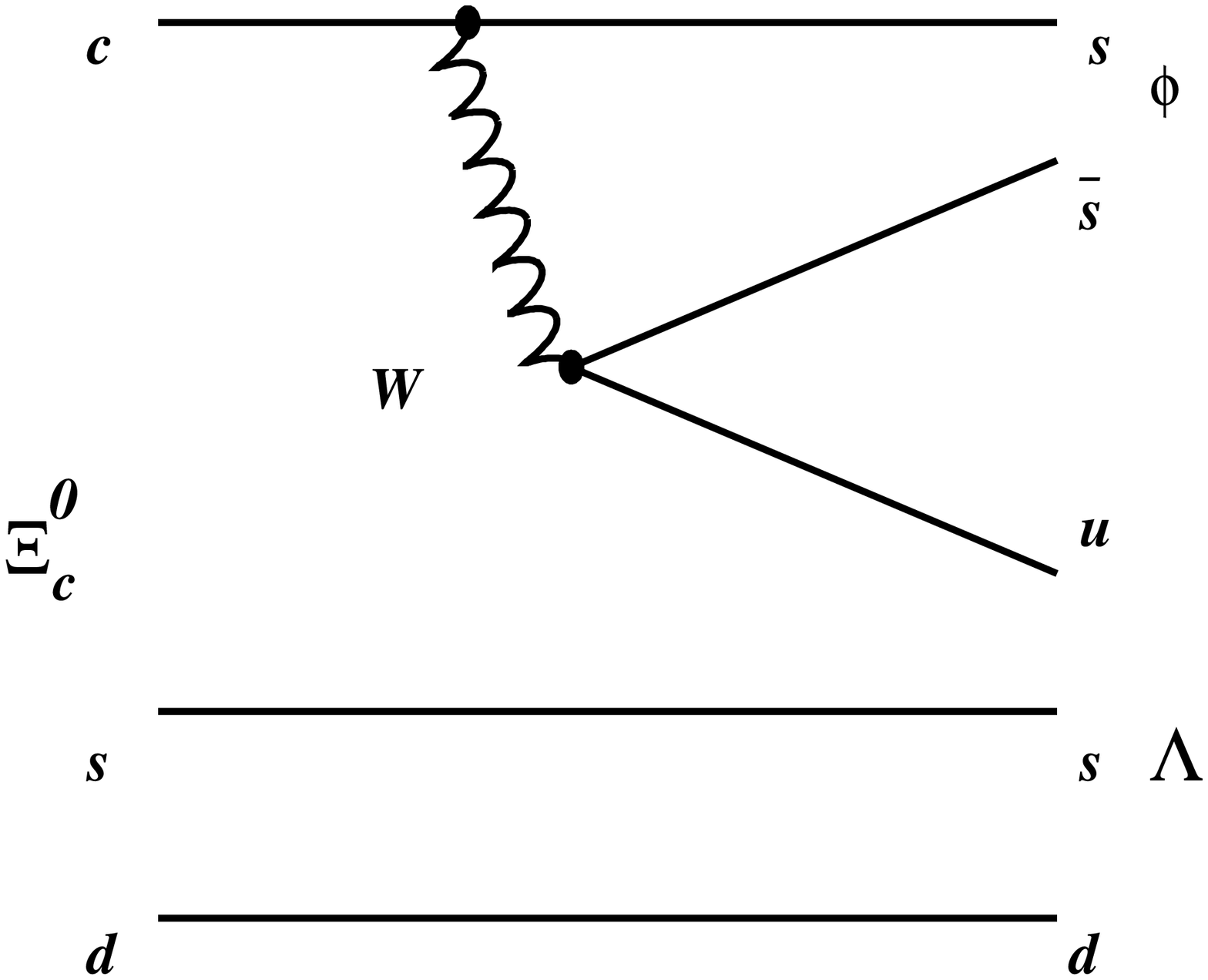}
\includegraphics[width=0.25\textwidth]{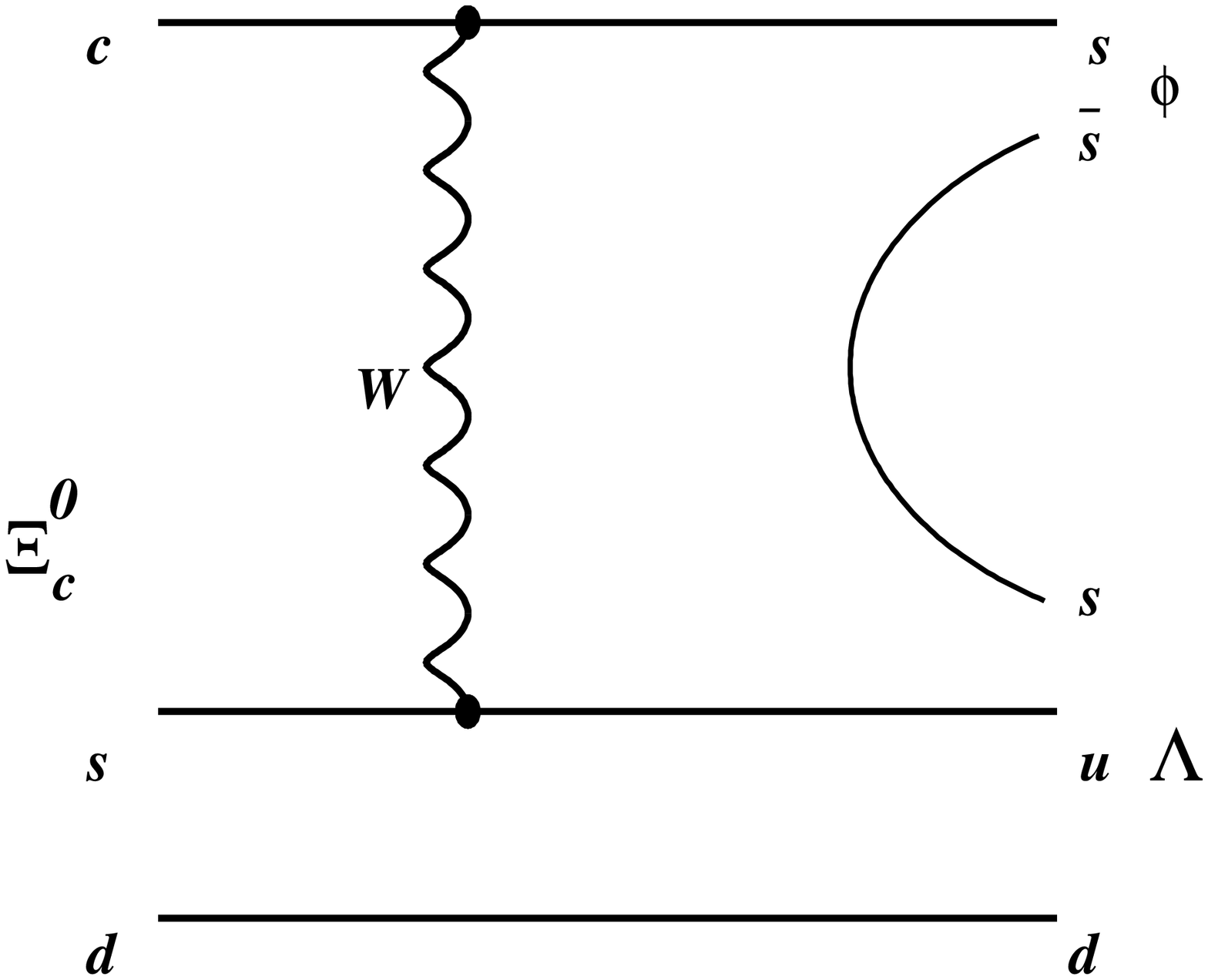}
 \caption{ 
Diagrams for the $\Xi_c^0\rightarrow \Xi^- K^+$ (upper two) and $\Xi_c^0\rightarrow  \Lambda \phi$ (lower two) decays. 
}
  \label{diagr}
\end{figure}

Weak decays of charmed baryons provide a useful test of many competing 
theoretical models
and approaches
~\cite{theoretical_predictions}. 
While many Cabibbo-favored decays of $\Lambda_c^+$ and $\Xi_c^{+,0}$ 
have been observed, the accuracy of the measured branching fractions 
remains poor.  
Some of the Cabibbo-suppressed (CS) decays of the $\Lambda_c^+$ 
were observed by Belle~\cite{vee2_belle_lamk} and {\it BABAR}~\cite{babar_lamc2lamk}; 
no experimental information 
is available for the CS decay modes of the $\Xi_c^0$. 
In this paper we report the first observation of the CS decays 
$\Xi_c^0\rightarrow \Xi^- K^+$, $\Xi_c^0\ra\Lambda K^+ K^-$ 
and $\Xi_c^0\rightarrow \Lambda \phi$.
The 
first 
decay is the Cabibbo-suppressed analogue 
of the Cabibbo-favored decay $\Xi_c^0\rightarrow \Xi^- \pi^+$ and proceeds 
through the external $W$-emission and $W$-exchange diagrams 
(see the two upper panels of Fig.~\ref{diagr}).
The 
third one 
proceeds through the internal 
$W$-emission and $W$-exchange diagrams (see lower two panels of 
Fig.~\ref{diagr}).
The $\Xi_c^0\ra\Lambda K^+ K^-$ decay can receive contributions 
from all $W$-mediated diagrams. 
The $W$-internal diagrams in charmed meson decays are usually 
color suppressed; this is not the case in charmed baryon decays~\cite{pdg}. 
Therefore, it is important to check 
this behavior in Cabibbo-suppressed $\Xi_c^0$ decays. 


The analysis is performed using data collected
with the Belle detector at the KEKB asymmetric-energy $e^+e^-$ collider~\cite{KEKB}.  
The data sample consists of $711$~fb$^{-1}$ taken at the $\Upsilon(4S)$ resonance.

The Belle detector is a large-solid-angle magnetic
spectrometer that consists of a silicon vertex detector,
a 50-layer central drift chamber (CDC), an array of
aerogel threshold Cherenkov counters (ACC),  
a barrel-like arrangement of time-of-flight
scintillation counters (TOF), and an electromagnetic calorimeter
composed of CsI(Tl) crystals located inside 
a superconducting solenoid coil that provides a 1.5~T
magnetic field.  An iron flux return located outside of
the coil is instrumented to detect $K_L^0$ mesons and to identify
muons.  The detector is described in detail elsewhere~\cite{Belle}.

We select charged pions, kaons and protons (unless a track has been 
identified as a daughter of  a 
$\Xi^-$ or $\Lambda$ hyperon) that originate from the region
$dr<0.5$~cm and $|dz|<1$~cm, where $dr$ and $dz$ are the distances
between the point of
closest approach and the interaction point (IP) in the plane
perpendicular to the beam axis (the r-$\phi$ plane) 
and along the beam direction ($z$), respectively.
%
%
We apply identification (ID) requirements for the 
charged particles using likelihoods $\mathcal{L}_K$, 
$\mathcal{L}_{\pi}$ and $\mathcal{L}_p$ for the kaon, pion and proton 
hypotheses, respectively, that are derived from information recorded
by the TOF, ACC and CDC.
Charged kaons are required to satisfy 
$\mathcal{L}_K/(\mathcal{L}_K+\mathcal{L}_{\pi})>0.6$ and 
$\mathcal{L}_K/(\mathcal{L}_K+\mathcal{L}_p)>0.6$. 
Protons are required to satisfy 
$\mathcal{L}_p/(\mathcal{L}_K+\mathcal{L}_p)>0.6$ and 
$\mathcal{L}_p/(\mathcal{L}_{\pi}+\mathcal{L}_p)>0.6$. 
For both species, these criteria have an efficiency greater than $87\%$ 
and a misidentification probability of less than $11\%$.
We apply no ID requirements for pions. 

In our Monte Carlo (MC) simulation, $\Xi_c^0$ baryons are produced in $e^+e^-\rightarrow c\bar{c}$ events 
 using the PYTHIA~\cite{simulation1} fragmentation package. 
Subsequent short-lived particle decays at the IP are generated by 
EvtGen~\cite{simulation1}. The detailed detector response is simulated
 using GEANT~\cite{simulation3}.  

The $\lam$ hyperons are reconstructed in the decay mode $\lam \rightarrow p \pi^-$.
(Unless stated otherwise, charge conjugation is implicitly assumed 
throughout the paper.) 
We fit the $p$ and $\pi^-$ tracks to a common vertex and  
require an invariant mass in a $\pm 3\,\mathrm{MeV}/{\it c}^2$ 
($\approx \pm 3\sigma$)
interval around the nominal $\Lambda$ mass. 
We then impose the following requirements on the $\lam$ decay vertex: 
  the vertex fit must be satisfactory; 
  the difference in the $z$ coordinates of the proton and
        pion at the decay vertex must satisfy $\Delta z<1$~cm; 
  the distance between the $\Lambda$ decay vertex position and IP
        in the $r$-$\phi$ plane must be greater than $0.1$~cm; 
   the angle $\alpha_{\Lambda}$ between the $\lam$ momentum vector
        and the vector joining the IP to the decay vertex
        must satisfy $\cos\alpha_{\Lambda} > 0.9$ for the case $\Xi_c^0\ra\Lambda \phi$. (No $\cos\alpha_{\Lambda}$ requirement is applied for the $\Xi_c^0\ra\Xi^- K^+$ candidates since, in this case,
 we select 
$\lam$'s emerging from the $\Xi^-$ decay vertex rather than the IP.)

The $\Xi^-$ hyperons are reconstructed in the decay mode $\Xi^-\ra \lam \pi^-$. 
We require a $\lam \pi^-$ invariant mass within a
$\pm 6$~MeV/{\it c}$^2$ 
($\approx \pm 3 \sigma$) 
interval around the nominal $\Xi^-$ mass,
fit the $\lam$ and the $\pi^-$ track to a common vertex and
apply the following requirements: 
the vertex fit must be satisfactory; 
the distance between the $\Xi^-$ decay vertex position and 
IP in the $r$-$\phi$ plane must be greater than $0.1$~cm;
the angle $\alpha_{\Xi^-}$ between the $\Xi^-$ momentum vector
        and the vector joining the IP to the $\Xi^-$ decay vertex
        must satisfy
       $\cos\alpha_{\Xi^-}>0.9$. 
All criteria described above and the reconstruction method for 
the two long-lived hyperons have been verified and used
in previous Belle papers on $\Lambda$, $\Xi^-$ and $\Omega^-$ 
hyperons~\cite{vee2_belle_lamk, vee2_belle_b2xiclamc, vee2_lesiak, vee2_belle_omegac}.  

The combinatorial background peaks at low momenta while charmed hadrons in 
$e^+e^-\ra c\bar{c}$ are concentrated at high momenta. 
Therefore, the momentum $p^*$ in the $e^+e^-$ center-of-mass frame 
for the $\Xi_c^0$ candidates is required to be greater than 
$3.0$~GeV/{\it c}. 



\begin{figure}[!h]
\includegraphics[width=0.45\textwidth]{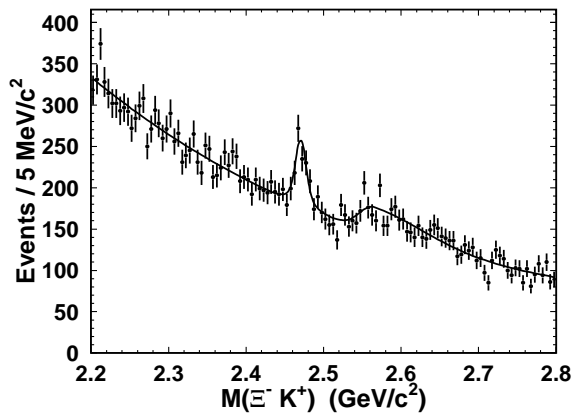}
 \caption{ Fitted $M(\Xi^- K^+)$ spectrum. The peak at 2470 MeV/{\it c}$^2$ 
corresponds to the $\Xi_c^0\ra\xim K^+$
signal. The broad structure from 2520~MeV/{\it c}$^2$ to 2700~MeV/{\it c}$^2$ corresponds to the  
$\Xi_c^0\ra\Xi^-\pi^+$. The smooth curve is the fit result, 
 described in the text. 
}
  \label{xic02xik_totfit}
\end{figure}

\begin{figure}[!h]
\includegraphics[width=0.45\textwidth]{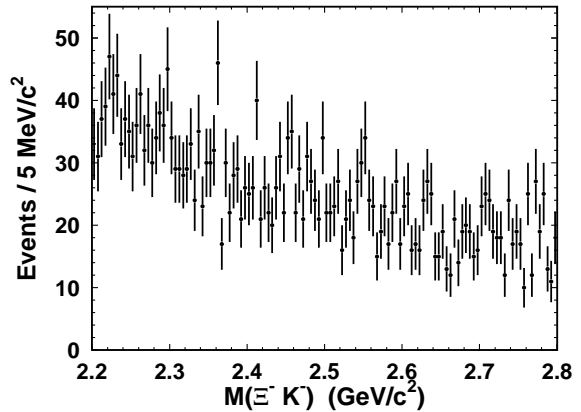}
 \caption{ The wrong-sign $M(\Xi^- K^-)$ spectrum. There are no peaking structures around 2470~MeV/{\it c}$^2$. 
}
  \label{xic02xik_ws}
\end{figure}

We reconstruct $\Xi_c^0\rightarrow \Xi^- K^+$  
candidates by combining $\Xi^-$ and $K^+$ candidates in the event. 
The resulting spectrum of the invariant mass $M(\Xi^- K^+)$ 
after all selection requirements 
 is shown in Fig.~\ref{xic02xik_totfit}, where a  
signal near 2470~MeV/{\it c}$^2$ is observed. 
In addition, a broad bump above the combinatorial background 
is evident at higher mass that 
 is due to a reflection from $\Xi_c^0\rightarrow \Xi^-\pi^+$,
in which the pion is misidentified as a kaon. 
We first check the origin of this reflection peak with data. 
With tight kaon ID requirements, the reflection bump completely vanishes; 
with looser 
ID requirements, the peak is more prominent. 
%
%
We check the shape and position of the reflection 
using signal MC events for the decay $\Xi_c^0\ra\Xi^-\pi^+$. 
By reconstructing 
such events as $\Xi^- K^+$, we observe that the position and shape 
of this reflection match those of the data. 
We also check the invariant mass distribution for 
the wrong-sign $\Xi^- K^-$ combinations in data with the same 
selection requirements. 
We find no indications of peaking structures and observe a  
mass distribution that is featureless over a wide mass range centered around the mass of the $\Xi_c^0$  
(see Fig.~\ref{xic02xik_ws}).

The solid curve in Fig.~\ref{xic02xik_totfit} is the result of the 
fit
that includes 
the signal, the reflection and the combinatorial background. 
Here and elsewhere in this paper, we use a binned maximum likelihood fit.   
The signal is described by a double Gaussian 
with a common floating mean and widths fixed from signal MC events.  
We calibrate these widths by the data-to-MC ratios 
from the study of $\Xi_c^0\ra\Xi^-\pi^+$ 
decay: we take $\sigma_{\rm core}$  and $\sigma_{\rm tail}$ 
from the fit to the $\Xi_c^0\ra\Xi^-\pi^+$ signal on  data 
and divide these by the corresponding $\sigma$'s from its signal MC events: 
$(\frac{\sigma_{\rm core}^{\rm data}}{\sigma_{\rm core}^{\rm mc}})_{\Xi^-\pi^+}=6.00/5.48=1.09$,
$(\frac{\sigma_{\rm tail}^{\rm data}}{\sigma_{\rm tail}^{\rm mc}})_{\Xi^-\pi^+}=12.50/11.06=1.13$. 
Then we make a correction of $\sigma_{\rm core}$ and $\sigma_{\rm tail}$ taken 
from $\Xi_c^0\ra\Xi^- K^+$ MC events 
and obtain the widths that we fix in the fit 
of the $\Xi_c^0\ra\Xi^-K^+$ signal on data:  
$(\sigma_{\rm core}^{\rm data})_{\Xi^- K^+}$=$1.09\times 5.87=6.43$~MeV/{\it c}$^2$,
$(\sigma_{\rm tail}^{\rm data})_{\Xi^- K^+}$=$1.13\times 12.72=14.35$~MeV/{\it c}$^2$.
We also include  
the shape of the reflection from $\Xi_c^0\ra\Xi^-\pi^+$ that is determined from
MC-generated $\Xi_c^0\ra\Xi^-\pi^+$ decays 
 reconstructed as $\Xi_c^0\ra\Xi^- K^+$. 
We find that the reflection can be parametrized by an asymmetric 
Gaussian with the right shoulder being larger than the left one.  
We fix the shape of the reflection
and leave its normalization as a free parameter in the fit.
The background is parametrized by a third-order polynomial function. 
The fit yields $N=313.8\pm 57.8$~events 
and $M=2470.6\pm 1.5$~MeV/{\it c}$^2$ for the $\Xi_c^0\ra\Xi^- K^+$ signal. 
The obtained mass is in good agreement with the world average 
mass of  $M(\Xi_c^0)=(2470.88_{-0.80}^{+0.34})$~MeV/{\it c}$^2$~\cite{pdg}.
The significance of the observed signal is   
8.0$\sigma$. 
The signal significance reported here and elsewhere in this paper 
is determined from 
2$\cdot {\rm ln}({\cal L}_0/{\cal L}_{\rm max})$, 
%
%
%
%
where ${\cal L}_{\rm max}$ is 
the maximum likelihood for the nominal fit and ${\cal L}_0$ 
is the corresponding value with the signal yield fixed to zero. 
The extraction of the significance takes into account 2 additional
 degrees of freedom (mass and yield).
%
%

We generate signal MC events without any momentum requirement, so here and elsewhere 
in this paper 
all calculated efficiencies take into account the kinematic efficiency of 
the $p^*>3.0$~GeV/{\it c} requirement. 
The measured total reconstruction efficiency for the 
$\Xi_c^0\ra \Xi^- K^+$ mode is $(4.47 \pm 0.03)\%$;
this includes the intermediate branching fraction 
${\cal B}(\Lambda\ra p\pi^-)$~\cite{pdg}.  
Using the results from the study of the normalization channel 
$\Xi^0_c\ra\Xi^-\pi^+$ 
(the number of events and reconstruction efficiency, described below), we 
obtain the ratio 
$\frac{{\cal B}(\Xi_c^0\ra \Xi^- K^+)}{{\cal B}(\Xi^0_c\ra\Xi^-\pi^+)}=(2.75\pm 0.51\pm 0.25)\times 10^{-2}$. 
The first and second errors are statistical and systematic, respectively.



\begin{figure}[!h]
\includegraphics[width=0.45\textwidth]{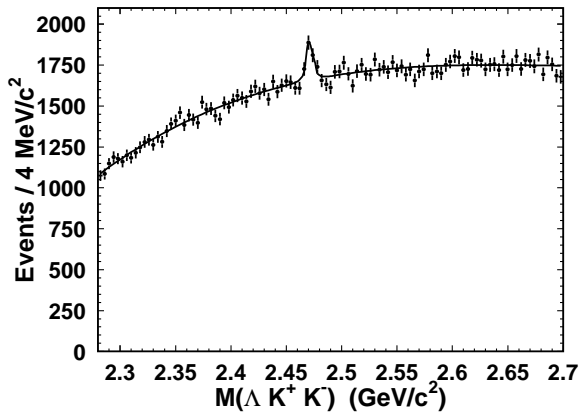}
\caption{The $M(\Lambda K^+ K^-)$ distribution together with the 
overlaid fitting curve. The fit is described in the text. }
\label{lkk_2gfit}
\end{figure}

In the search for $\Xi_c^0\rightarrow \Lambda \phi (\phi\ra K^+ K^-)$ decay,  
in addition to the above-described selection criteria, we require that 
the mass of the $\Lambda K^-$ pair 
be outside 
a $\pm 6.5$~MeV/{\it c}$^2$ mass 
window around
$M(\Omega^-)=1672.45$~MeV/{\it c}$^2$~\cite{pdg}.  
This requirement removes the contribution from 
the well-known $\Xi_c^0\ra\Omega^- K^+$ ($\Omega^-\ra\Lambda K^-$) decay~\cite{pdg}. 
The resulting spectrum of the three-body invariant mass
 $M(\Lambda K^+ K^-)$  
is shown in Fig.~\ref{lkk_2gfit}, where  a 
signal near 2470~MeV/{\it c}$^2$ is observed. 
We fit the  $\Xi_c^0\rightarrow \Lambda K^+ K^-$   signal to the data 
with a double Gaussian with the fixed widths from corresponding MC events 
($\sigma_{\rm core}=2.53$~MeV/{\it c}$^2$, $\sigma_{\rm tail}=6.10$~MeV/{\it c}$^2$). 
For the background, we use a third-order polynomial. 
The fit results in a mass of $M=2471.2\pm 1.1$~MeV/{\it c}$^2$   
and a yield of $N=511.0\pm 109.5$. 
This mass is in good agreement with the world average 
mass of the $\Xi_c^0$.
The significance 
of this signal is 
$6.4\sigma$.

\begin{figure}[h]
\includegraphics[width=0.45\textwidth]{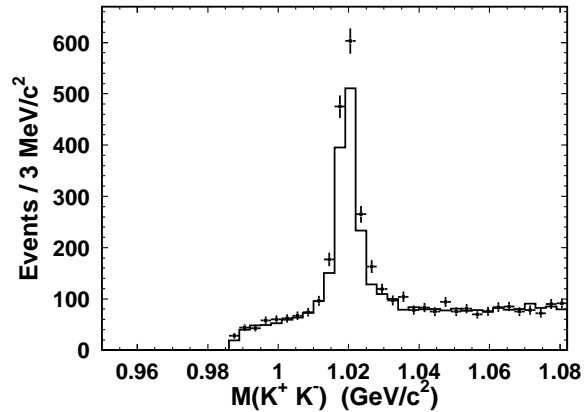}
 \caption{
The $M(K^+ K^-)$ distributions. Cross points with error bars represent
 the events within 
the $\pm 12$~MeV/{\it c}$^2$ mass window around $M(\Lambda K^+ K^-)=2471.2$~MeV/{\it c}$^2$. 
The solid histogram shows the $\phi$ signal 
for the events taken from the normalized $\Xi_c^0$ sidebands (see the text).
}
  \label{mkk}
\end{figure}

To obtain the $\Xi_c^0\rightarrow \Lambda \phi$ signal, 
we select $\Lambda K^+ K^-$ combinations within 
the $\pm 12$~MeV/{\it c}$^2$ mass window around $M(\Xi_c^0)=2470.9$~MeV/{\it c}$^2$
and 
investigate the distribution of 
$M(K^+ K^-)$ 
shown by the data points in 
Fig.~\ref{mkk}. 
The superimposed histogram 
 shows 
the $\phi$ signal for the events taken from the $\Xi_c^0$ sidebands, 
which are normalized to the area under the $\Xi_c^0$ signal.
The left $\Xi_c^0$ sideband is defined by 
$2403.2$~MeV/{\it c}$^2$$<M(\Lambda K^+ K^-)<2451.2$~MeV/{\it c}$^2$, 
and the right one by  
$2491.2$~MeV/{\it c}$^2$$<M(\Lambda K^+ K^-)<2539.2$~MeV/{\it c}$^2$.
A distinct excess of $\phi$ mesons is observed in the $\Xi_c^0$ signal region, 
establishing the observation of the two-body $\Xi_c^0\ra\Lambda\phi$ decay.

The $\phi$ signal is described by a 
Breit-Wigner function convolved with the double Gaussian resolution function 
with the widths fixed from MC events 
($\sigma_{\rm core}=0.61$~MeV/{\it c}$^2$, $\sigma_{\rm tail}=1.39$~MeV/{\it c}$^2$).
The natural width $\Gamma_{\phi}$ is fixed to its nominal value of 
4.26~MeV~\cite{pdg}. 
The threshold function multiplied 
by a third-order polynomial is used to model the combinatorial background 
together with a nonresonant contribution. 
%
%
%
%
The fit results in the following $\phi$ yields: 
$N_{1}=1533.1\pm 47.9$ events in the $\Xi_c^0$ signal region and 
$N_{2}=5006.8\pm 88.8$ events in the $\Xi_c^0$ sidebands region. 
From this, the final 
net $\phi$ yield in $\Xi_c^0\rightarrow \Lambda K^+K^-$ 
decays is $N_{\phi} = (N_{1}\pm \delta N_{1})/0.98-(N_{2}\pm \delta N_{2})\times 0.249 =315.8\pm 53.7$.  
The coefficient 0.98 takes into account the efficiency of the 
mass requirement of $\pm 12$~MeV/{\it c}$^2$ around $M(\Xi_c^0)$. The coefficient 0.249 
is the ratio of 
areas under the $\Xi_c^0$ signal and the sum of its sidebands. 
%
%
%
%
From the obtained $\phi$ net yield and the 
probability value of the Gaussian 
distribution of the error, we extract a significance of $5.9\sigma$
for the 
$\Xi_c^0\rightarrow\Lambda\phi$ signal.
By varying the width of the $\Xi_c^0$ sidebands and repeating the $\phi$ 
yield extraction procedure, we obtain significances that are  
never less than $5.6\sigma$. We quote this latter value 
as our significance of the $\Xi_c^0\rightarrow\Lambda\phi$ signal, 
including the systematic error.   
%
%
%
The total reconstruction efficiency, including the 
intermediate branching fractions of $\Lambda\ra p\pi^-$ and $\phi\ra K^+K^-$ 
is extracted from signal MC events to be $(3.60\pm 0.02)\%$. 
We obtain 
$\frac{{\cal B}(\Xi_c^0\ra\Lambda \phi)}{{\cal B}(\Xi_c^0\rightarrow\Xi^- \pi^+)}=(3.43\pm 0.58\pm 0.32)\times 10^{-2}$.
The first and second errors are statistical and systematic, respectively.

To obtain the ratio of branching fractions for the 
  three-body $\Xi_c^0\rightarrow\Lambda K^+ K^-$ channel, 
we estimate its signal efficiency as follows. 
Taking into account the correspondence between 
the obtained number of events for the three-body mode 
($511.0\pm 109.5$) and for the  
$\Xi_c^0\rightarrow\Lambda\phi$ mode ($315.8\pm 53.7$), 
we generate a sample of $\Xi_c^0$ states that decay
$40\%$ of the time into the three-body phase space 
$\Lambda K^+ K^-$ final state and 
 $60\%$ of the time into the $\Lambda\phi$ final state. 
The total reconstruction efficiency, including 
the intermediate branching fraction for $\Lambda\ra p\pi^-$,  
is found to be $(7.01\pm 0.04)\%$.
We vary the portion of the resonant mode over a 
$\pm 50\%$ range 
and repeat the efficiency extraction. 
The absolute value of the largest variation in the 
total reconstruction efficiency is found to be $0.15\%$, 
which is treated as a systematic error. 
Finally, 
we get 
$\frac{{\cal B}(\Xi_c^0\ra\Lambda K^+ K^-)}{{\cal B}(\Xi_c^0\rightarrow\Xi^- \pi^+)}=(2.86\pm 0.61\pm 0.37)\times 10^{-2}$, 
where the first and second errors are statistical and systematic, respectively.
An additional source of systematic error due to the MC model 
of $\Xi_c^0$ decay into the final state $\Lambda K^+ K^-$ 
is also included.



\begin{figure}[!h]
\includegraphics[width=0.45\textwidth]{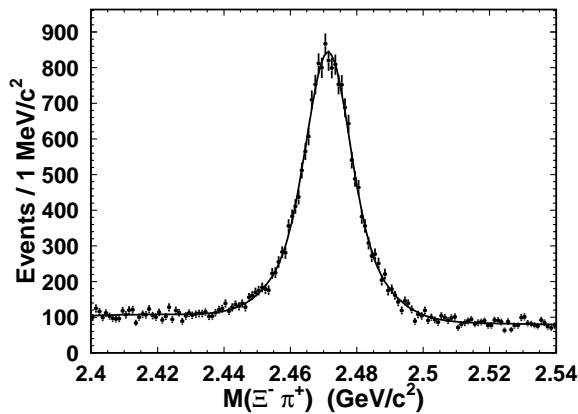}
 \caption{ 
The $M(\Xi^- \pi^+)$ distribution for data together with the 
overlaid fitting curve. 
The fit is described in the text.}
  \label{xic02xipi}
\end{figure}

Currently, there are no absolute branching fraction measurements for $\Xi_c^0$, 
so we choose to normalize 
the results for $\Xi_c^0$ decays 
to the well-known decay mode 
$\Xi^0_c\ra\Xi^-\pi^+$. 
Using the same data sample, the selection criteria and the 
$p(\Xi_c^0)^*>3.0$~GeV/{\it c} requirement 
described above,  
we reconstruct $\Xi^0_c\ra\Xi^-\pi^+$ decays and obtain 
the $M(\Xi^-\pi^+)$ spectrum shown in Fig.~\ref{xic02xipi}. 
We fit this spectrum with a double Gaussian with a floating common mean 
and floating widths
(to describe the signal) and a third-order polynomial function 
to account for the background. 
The signal yield is $N=15324\pm 262$.  
The Gaussian widths and common mean are extracted from the fit to be 
$\sigma_{\rm core}=6.0\pm 0.3$~MeV/{\it c}$^2$, $\sigma{\rm tail}=12.5\pm 0.8$~MeV/{\it c}$^2$ and $M=2471.4\pm
0.1$~MeV/{\it c}$^2$, respectively. 
The mass is in agreement with the 
world average value: $M(\Xi_c^0)=(2470.88_{-0.80}^{+0.34})$~MeV/{\it c}$^2$~\cite{pdg}. 
We generate signal MC events and reconstruct 
the generated events according to the procedure that is used 
in analyzing the data. 
The total reconstruction efficiency 
is determined to be $(6.00 \pm 0.03)\%$.   
This efficiency includes the intermediate branching fraction 
${\cal B}(\Lambda\ra p\pi^-)$~\cite{pdg}.  
Since the number of signal events for this mode is large, we do not fix the  
Gaussian resolution to obtain the final yield 
for $\Xi_c^0\ra\Xi^-\pi^+$. 

We consider the following sources of systematic errors: 
the fit,
$K$ ID efficiency and 
MC statistics.
The fit systematics are determined by varying the range 
of the fitted invariant mass distributions
and by changing the polynomial order for the background function. 
Other sources of uncertainties, such as  
particle reconstruction efficiency and $\Lambda$ 
reconstruction efficiency, cancel in the branching fraction ratio. 
For the $\Xi_c^0\ra\Lambda\phi$ and   $\Xi_c^0\ra\Lambda K^+ K^-$ 
results, we consider possible 
interference between the non-$\phi$ $\Lambda K^+ K^-$ and resonant 
$\Lambda \phi(\phi\ra K^+ K^-)$ amplitudes. This effect is estimated 
to be $3.8\%$. 
Finally, the MC model of the   $\Xi_c^0\ra\Lambda K^+ K^-$ 
mode introduces an additional uncertainty that we estimate to be $2\%$. 
As we do not have a calibration channel for
the width correction in the $\Xi_c^0\ra \Lambda K^+K^-$ mode, we add a $10\%$
systematic error based on the calculated corrections in the $\Xi_c^0 \ra \Xi^- \pi^+$ mode.
Table~\ref{systematics1} 
summarizes the systematic errors. 


\begin{table*}[!b]
\begin{ruledtabular}
\caption {Summary of systematic errors in the ratios of $\frac{{\cal B}(\Xi_c^0\rightarrow\Xi^- K^+)}{{\cal B}(\Xi_c^0\rightarrow\Xi^- \pi^+)}$ ($\Xi^- K^+$), $\frac{{\cal B}(\Xi_c^0\ra\Lambda K^+ K^-)}{{\cal B}(\Xi_c^0\rightarrow\Xi^- \pi^+)}$ ($\Lambda K^+ K^-$) and $\frac{{\cal B}(\Xi_c^0\ra\Lambda \phi)}{{\cal B}(\Xi_c^0\rightarrow\Xi^- \pi^+)}$ ($\Lambda \phi$).}
\begin{tabular}{c|c|c|c}
Source  & Value, $\%$ ($\Xi^- K^+$) &   Value, $\%$ ($\Lambda \phi$) & Value, $\%$ ($\Lambda K^+ K^-$)\\
\hline
Kaon ID  & 1 & 2 & 2\\
Fit model   & 9 & 8 & 7\\
Interference & ... & 3.8 & 3.8 \\
MC statistics  & 0.5 & 0.5 & 0.5\\
MC model & ... & ... & 2 \\
MC width & ... & ... & 10 \\
\hline
Total & 9.1 & 9.1 & 13.1\\
\end{tabular}
\label{systematics1}
\end{ruledtabular}
\end{table*}

In conclusion, we have  
observed for the first time the 
Cabibbo-suppressed decays
  $\Xi_c^0\rightarrow\Xi^- K^+$, $\Xi_c^0\ra\Lambda K^+ K^-$  
and $\Xi_c^0\ra\Lambda\phi$ with significances of
8.0$\sigma$, 6.4$\sigma$ and 5.6$\sigma$, respectively. 
The ratios of the branching fractions 
$\frac{{\cal B}(\Xi_c^0\rightarrow\Xi^- K^+)}{{\cal B}(\Xi_c^0\rightarrow\Xi^- \pi^+)}$, 
$\frac{{\cal B}(\Xi_c^0\ra\Lambda K^+ K^-)}{{\cal B}(\Xi_c^0\rightarrow\Xi^- \pi^+)}$ 
and $\frac{{\cal B}(\Xi_c^0\ra\Lambda\phi)}{{\cal B}(\Xi_c^0\rightarrow\Xi^- \pi^+)}$ 
are measured to be $(2.75\pm 0.51\pm 0.25)\times 10^{-2}$,  
$(2.86\pm 0.61\pm 0.37)\times 10^{-2}$ 
and $(3.43\pm 0.58\pm 0.32)\times 10^{-2}$, respectively.   

The observed decay modes proceed through 
external and internal $W$-emission diagrams with an admixture 
of the $W$-exchange diagram. 
Our results can 
be used  
to study the corresponding decay dynamics and to investigate quantitatively
the interplay between strong and weak interactions
in charmed baryon weak decays.  
We confirm the previous observations~\cite{vee2_belle_lamk, babar_lamc2lamk, babar_xic02omk} 
that the $W$-internal diagrams are not (color) suppressed  
as compared to 
the $W$-external diagrams in charm baryon decays. 


We thank the KEKB group for the excellent operation of the
accelerator; the KEK cryogenics group for the efficient
operation of the solenoid; and the KEK computer group,
the National Institute of Informatics, and the 
PNNL/EMSL computing group for valuable computing
and SINET4 network support.  We acknowledge support from
the Ministry of Education, Culture, Sports, Science, and
Technology (MEXT) of Japan, the Japan Society for the 
Promotion of Science (JSPS), and the Tau-Lepton Physics 
Research Center of Nagoya University; 
the Australian Research Council and the Australian 
Department of Industry, Innovation, Science and Research;
the National Natural Science Foundation of China under
Contracts No.~10575109, No.~10775142, No.~10875115 and No.~10825524; 
the Ministry of Education, Youth and Sports of the Czech 
Republic under Contracts No.~LA10033 and MSM0021620859;
the Department of Science and Technology of India; 
the Istituto Nazionale di Fisica Nucleare of Italy; 
the BK21 and WCU program of the Ministry Education Science and
Technology, National Research Foundation of Korea,
and GSDC of the Korea Institute of Science and Technology Information;
the Polish Ministry of Science and Higher Education;
the Ministry of Education and Science of the Russian
Federation and the Russian Federal Agency for Atomic Energy;
the Slovenian Research Agency;  the Swiss
National Science Foundation; the National Science Council
and the Ministry of Education of Taiwan; and the U.S.\
Department of Energy and the National Science Foundation.
This work is supported by a Grant-in-Aid from MEXT for 
Science Research in a Priority Area (``New Development of 
Flavor Physics''), and from JSPS for Creative Scientific 
Research (``Evolution of Tau-lepton Physics'').

\end{document}